\documentclass[pra, twocolumn, showpacs, showkeys]{revtex4}

\usepackage{amsmath}
\usepackage{amssymb}
\usepackage[dvips]{graphicx}

\DeclareGraphicsExtensions{.ps}

\newcommand*{\etal}{\textit{et al.\ }}

\begin{document}
\title{Reply to 'Comment on ``Detuning effects in the one-photon mazer'' '}
\date{\today}
\author{Thierry Bastin}
\email{T.Bastin@ulg.ac.be}
\author{John Martin}
\email{John.Martin@ulg.ac.be} \affiliation{Institut de Physique
Nucl\'eaire, Atomique et de Spectroscopie, Universit\'e de Li\`ege
au Sart Tilman, B\^at.\ B15, B - 4000 Li\`ege, Belgique}

\begin{abstract}
We refute in this Reply the criticisms made by M. Abdel-Aty. We
show that none of them are founded and we demonstrate very
explicitly what is wrong in the arguments developed by this
author.
\end{abstract}



\maketitle

\begin{figure}
\begin{center}
\noindent\mbox{\includegraphics[width=8cm, bb= 60 615 580
805]{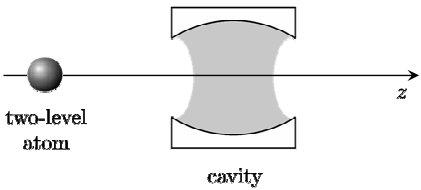}}
\end{center}
\caption{General scheme of the mazer.} \label{mazerfig}
\end{figure}

In a recent paper~\cite{Bas03b} we have presented the quantum
theory of the mazer in the nonresonant case, pointing out various
new effects compared to the resonant case. In this system, the
interaction of a two-level atom with a single mode of a cavity is
investigated. The atom is supposed to move unidirectionally on the
way to the cavity and the interaction occurs when the atom passes
through it (see Fig.~\ref{mazerfig}). The effects of this
interaction are then studied after the atom has left the cavity
region. Compared to the conventional micromaser, the atomic motion
is described quantum mechanically (see our paper~\cite{Bas03b} and
references therein for the detailed description of the system and
the model considered). In the nonresonant case, a detuning between
the cavity mode and the atomic transition frequencies is present.
In our paper, we showed that this detuning adds a potential step
effect not present at resonance, resulting in a well-defined
acceleration or deceleration (according to the sign of the
detuning) of the atoms that emit a photon inside the cavity if
they are initially excited. We also demonstrated that this photon
emission inside the cavity may be completely blocked by use of a
positive detuning. Finally, we characterized the properties of the
induced emission probability in various regimes and demonstrated
notably that the well-known Rabi formula is well recovered by the
general quantum theory in the hot atom regime (where the
quantization of the center-of-mass motion is not necessary).
Various criticisms about our paper have been raised by Abdel-Aty
and summarized in a Comment to which this Reply is intended. We
give here an answer to all of these criticisms and demonstrate
that none of them are founded.

Firstly, and contrary to what is claimed in Abdel-Aty's Comment,
it is obvious that all the physical effects just described and the
way they have been derived in Ref.~\cite{Bas03b} do not appear in
any form in any previous paper dedicated to this subject by this
author, namely, Refs.~\cite{Abd02a} and~\cite{Abd02b}. This will
be further demonstrated in the rest of this Reply.

Secondly, it is stated by Abdel-Aty that the evaluation of the
coupled equations (5a) and (5b) of our paper is not satisfactory.
In particular, it is declared that, in the first line of Eq.~(5a)
(resp.~(5b)), $\cos^2 \theta$ (resp.~$\sin^2 \theta$) should be
replaced by $\cos 2\theta$ (resp.~$\sin 2\theta$). We wonder about
this criticism as it is in no way argued and actually is wrong. To
be very explicit, we give hereafter the details of the
calculations which lead to Eqs. (5a) and (5b) and we demonstrate
that they are perfectly correct.

The Hamiltonian used to describe the mazer is given
by~\cite{Mey97}
\begin{equation}
    \label{Hamiltonian}
        H = \hbar \omega_0 \sigma^{\dagger} \sigma + \hbar \omega a^{\dagger} a + \frac{p^2}{2m}+ \hbar g \, u(z) (a^{\dagger} \sigma + a
        \sigma^{\dagger}),
\end{equation}
with usual notations. In particular, $\omega$ and $\omega_0$ are
the frequencies of the cavity mode and the atomic transition,
respectively. The atomic motion is defined along the $z$ direction
and $u(z)$ describes the spatial variation of the atom-cavity
coupling (the so-called cavity mode function). The two atomic
internal states are denoted by $|a\rangle$ (excited state) and
$|b\rangle$ (ground state). The global Hilbert space of the system
is given by
\begin{equation}
    \label{HilbertSpace}
    \mathcal{H} = \mathcal{E}_z \otimes \mathcal{E}_A \otimes
    \mathcal{E}_R
\end{equation}
with $\mathcal{E}_z$ the space of the wave functions describing
the one-dimensional atomic center-of-mass motion, $\mathcal{E}_A$
the space describing the atomic internal degree of freedom, and
$\mathcal{E}_R$ the space of the cavity single mode radiation.

We introduce in the space $\mathcal{E}_A \otimes \mathcal{E}_R$ the
orthonormal basis
\begin{equation}
\label{basis} \left\{ \begin{array}{l}
|\Gamma_{-1}\rangle = |b,0\rangle, \\
|\Gamma_{n}^+(\theta)\rangle = \cos\theta\,|a,n\rangle +
\sin\theta\,|b,n+1\rangle,\\
|\Gamma_{n}^-(\theta)\rangle = -\sin\theta\,|a,n\rangle +
\cos\theta\,|b,n+1\rangle,
\end{array} \right.
\end{equation}
with $\theta$ an arbitrary parameter and $n \geqslant 0$. Combined
with the $z$ representation $\{|z\rangle\}$ in $\mathcal{E}_z$,
the set of vectors
\begin{equation}
\label{globalBasis}
\{|z,\Gamma_{-1}\rangle, |z,\Gamma_{n}^{\pm}(\theta)\rangle\}
\end{equation}
defines an orthonormal basis over the whole Hilbert space~$\mathcal{H}$.

Projecting the Schr\"odinger equation
\begin{equation}
i \hbar \frac{d}{dt} |\psi(t)\rangle = H |\psi(t)\rangle
\end{equation}
onto the basis~(\ref{globalBasis}) yields for every $n \geqslant 0$
\begin{equation}
\label{seq}
i \hbar \frac{\partial}{\partial t} \psi_{n,\theta}^{\pm}(z,t) = \langle z, \Gamma_{n}^{\pm}(\theta)| H |\psi(t)\rangle
\end{equation}
with
\begin{equation}
\psi_{n,\theta}^{\pm}(z,t) \equiv \langle z, \Gamma_{n}^{\pm}(\theta) |\psi(t)\rangle
\end{equation}

Using the completeness relation
\begin{align}
    1 = & \int dz \Big[ \sum_{n=0}^{\infty} \left( |z, \Gamma_{n}^+(\theta)\rangle \langle z, \Gamma_{n}^+(\theta)| + |z, \Gamma_{n}^-(\theta)\rangle \langle z, \Gamma_{n}^-(\theta)|\right) \nonumber \\
    &  \qquad \qquad + |z, \Gamma_{-1}\rangle \langle z, \Gamma_{-1}| \Big],
\end{align}
the right-hand side of Eq.~(\ref{seq}) reads
\begin{widetext}
\begin{align}
\label{rt} \langle z, \Gamma_{n}^{\pm}(\theta)| H |\psi(t)\rangle
= \int dz' \Big[ & \sum_{n'=0}^{\infty} \big( \langle z,
\Gamma_{n}^{\pm}(\theta)| H |z', \Gamma_{n'}^+(\theta)\rangle
\psi_{n',\theta}^{+}(z',t)
+ \langle z, \Gamma_{n}^{\pm}(\theta)| H |z', \Gamma_{n'}^-(\theta)\rangle \psi_{n',\theta}^{-}(z',t) \big) \nonumber \\
& + \langle z, \Gamma_{n}^{\pm}(\theta)| H |z', \Gamma_{-1} \rangle \langle z', \Gamma_{-1}| \psi(t) \rangle \Big].
\end{align}
\end{widetext}

Straightforward calculations yield
\begin{align}
\langle \Gamma_{n}^{+}(\theta) | \sigma^{\dagger} \sigma |
\Gamma_{n'}^{+}(\theta) \rangle & = \cos^2 \theta \, \delta_{nn'},
\\
\langle \Gamma_{n}^{+}(\theta) | \sigma^{\dagger} \sigma |
\Gamma_{n'}^{-}(\theta) \rangle & = -\frac{1}{2}\sin2\theta \,
\delta_{nn'},
\end{align}
\begin{align}
\langle \Gamma_{n}^{+}(\theta) | a^{\dagger} a |
\Gamma_{n'}^{+}(\theta) \rangle & = (n + \sin^2 \theta) \,
\delta_{nn'}, \\
\langle \Gamma_{n}^{+}(\theta) | a^{\dagger} a |
\Gamma_{n'}^{-}(\theta) \rangle & = \frac{1}{2}\sin2\theta \,
\delta_{nn'},
\end{align}
\begin{align}
\langle \Gamma_{n}^{+}(\theta) | a^{\dagger} \sigma + a
\sigma^{\dagger} | \Gamma_{n'}^{+}(\theta) \rangle & = \sqrt{n+1}
\, \sin 2 \theta \, \delta_{nn'},
\\
\langle \Gamma_{n}^{+}(\theta) | a^{\dagger} \sigma + a
\sigma^{\dagger} | \Gamma_{n'}^{-}(\theta) \rangle & = \sqrt{n+1}
\, \cos 2 \theta \, \delta_{nn'}
\end{align}
and
\begin{equation}
\langle z, \Gamma_{n}^{+}(\theta)| H |z', \Gamma_{-1} \rangle = 0.
\end{equation}

We thus have from Eq.~(\ref{rt})
\begin{widetext}
\begin{align}
\label{detail1} \langle z, \Gamma_{n}^{+}(\theta)| H
|\psi(t)\rangle = \int dz' \Big[ & \sum_{n'=0}^{\infty}
\delta_{nn'} \delta(z-z') \left( \frac{-\hbar^2}{2m}
\frac{\partial^2}{\partial z'^2} + \hbar \omega_0 \cos^2 \theta + \hbar \omega (n + \sin^2 \theta) + \hbar g \sqrt{n+1} \, u(z') \sin 2 \theta \right)\psi_{n',\theta}^{+}(z',t) \nonumber \\
& + \delta_{nn'} \delta(z-z')\left(-\frac{\hbar \omega_0}{2} \sin
2 \theta + \frac{\hbar \omega}{2} \sin 2 \theta + \hbar g
\sqrt{n+1} \, u(z') \cos 2 \theta
\right)\psi_{n',\theta}^{-}(z',t) \Big].
\end{align}
\end{widetext}

Defining the detuning $\delta = \omega - \omega_0$ and inserting
Eq.~(\ref{detail1}) into Eq.~(\ref{seq}) yields the Schr\"odinger
equation for the $\psi^+_{n, \theta}(z,t)$ component, that is
\begin{align}
i\hbar\frac{\partial}{\partial t}\psi^+_{n, \theta}(z,t)= & \bigg[
  -\frac{\hbar^2}{2m}\frac{\partial^2}{\partial
    z^2}+(n+1)\hbar\omega-\cos^2\theta\:\hbar\delta\; \nonumber \\
  & +\;\hbar g
  u(z) \sqrt{n\!+\!1}\sin2\theta\bigg]\psi^+_{n, \theta}(z,t) \nonumber \\
& +\bigg[\hbar g u(z)
\sqrt{n\!+\!1}\;\cos2\theta \nonumber \\
& +\frac{1}{2}\sin2\theta\:\hbar\delta\bigg]\psi^-_{n,
\theta}(z,t),
\end{align}
which is exactly Eq.~(5a) of our paper~\cite{Bas03b} (where of
course $\frac{1}{2} \theta \hbar \delta \sin 2$ must be read
$\frac{1}{2} \hbar \delta \sin 2 \theta$. This glaring typographic
error was introduced after proof corrections and is beyond the
scope of this Comment). Therefore and contrary to what is claimed
by Abdel-Aty, the $\cos^2 \theta$ term in the first line of our
equation (5a) in Ref.~\cite{Bas03b} \emph{must not} be replaced by
$\cos 2 \theta$, confirming that this equation is perfectly
correct.

We demonstrate in a similar manner that Eq.~(5b) of our paper
(which yields the Schr\"odinger equation for the $\psi^-_{n,
\theta}(z,t)$ component) is also error free and that again the
$\sin^2 \theta$ term that appears in the first line of this
equation \emph{must not} be replaced by $\sin 2 \theta$. We
therefore refute the criticism made in Abdel-Aty's Comment that
these equations would not be satisfactory. They are, and there is
absolutely no need to replace them with those proposed by this
author, namely Eq.~(9) in his Comment. We demonstrate even at the
end of this Reply that this replacement cannot be done as it leads
to inconsistencies and wrong results in some cases, especially in
the mesa mode case that is presently investigated by us and
Abdel-Aty.

A third criticism states that we would have overlooked the
formulae for $\cos 2 \theta_n$ and $\sin 2\theta_n$ where
$\theta_n$ defines the dressed-state basis and is given by
\begin{equation}
    \label{thetads}
    \cot 2 \theta_n = - \frac{\delta}{\Omega_n}\, ,
\end{equation}
with $\Omega_n = 2 g \sqrt{n + 1}$. According to this criticism,
we would have avoided great simplifications of our equations.
Nothing is more wrong. We did not overlook anything. It is well
explained in our paper that, in the case of the mesa mode function
($u(z) = 1$ for $0 < z < L$, 0 elsewhere~\cite{Eng91}, where $L$
is the length of the cavity supposed to be located in the $[0,L]$
region), the coupled equations (5a) and (5b) reduce to a much
simpler decoupled form in the local dressed state basis, namely,
Eqs.~(7) and (17) of our paper (for outside and inside the cavity,
respectively). Equation (20) of our paper does not express anymore
than Eq.~(10) of Abdel-Aty's Comment. Indeed, according to our
notations, Eq.~(20) of Ref.~\cite{Bas03b} yields explicitly
\begin{align}
\cos 2\theta_n & = 1 - 2 \sin^2 \theta_n = 1 - \frac{\Lambda_n + \delta}{\Lambda_n} = -\frac{\delta}{\Lambda_n}, \\
\sin 2\theta_n & = 2 \sin\theta_n \cos\theta_n = \frac{\sqrt{\Lambda_n^2 - \delta^2}}{\Lambda_n} = \frac{\Omega_n}{\Lambda_n},
\end{align}
with $\Lambda_n = \sqrt{\delta^2 + \Omega_n^2}$.

Finally, we prove now that Eq.~(9) in Abdel-Aty's Comment does not
express another point of view equivalent to our equations and
that, contrary to the claim of this author, it cannot replace
Eqs.~(5a) and (5b) of our paper.

Their equations read (taking into account that the first partial
derivatives on the left-hand side of Eqs.~(8) and (9) in
Abdel-Aty's Comment must evidently be read $\partial/\partial t$
instead of $\partial/\partial z$ !)
\begin{align}
\label{aeq}
i \frac{\partial C_n^+}{\partial t} = & \left( -\frac{1}{2M}\frac{\partial^2}{\partial z^2} + V_n^+ - \left( \frac{d\theta_n}{dz} \right)^2 \right) C_n^+ \nonumber \\
& - \left( 2 \frac{\partial C_n^-}{\partial z} \left( \frac{d\theta_n}{dz}\right) + C_n^- \left( \frac{d\theta_n}{dz}\right)^2 \right), \nonumber \\
i \frac{\partial C_n^-}{\partial t} = & -\left( -\frac{1}{2M}\frac{\partial^2}{\partial z^2} + V_n^- - \left( \frac{d\theta_n}{dz} \right)^2 \right) C_n^- \nonumber \\
& + \left( 2 \frac{\partial C_n^+}{\partial z} \left( \frac{d\theta_n}{dz} \right) + C_n^+ \left( \frac{d\theta_n}{dz}\right)^2 \right).
\end{align}
using the notations as defined in the Comment, where $C^{\pm}_n$
are the components of the wave function over the dressed states,
the $\theta_n$ angle is $z$ dependent through the relation
\begin{equation}
\label{thetan}
\cot 2 \theta_n = - \frac{\delta}{2 g u(z) \sqrt{n+ 1}}
\end{equation}
and where
\begin{equation}
    \label{pot}
    V_n^{\pm} = \left( n + \frac{1}{2} \right) \omega \pm \frac{1}{2}\sqrt{\delta^2 + 4 g^2 u^2(z) (n+1)}\,.
\end{equation}

At resonance ($\delta = 0$), the $\theta_n$ angle that defines the
dressed state basis takes the value $\pi/4$ everywhere. Thus
$\frac{d\theta_n}{dz}~=~0$ whatever the cavity mode function and
Eq.~(\ref{aeq}) must reduce to the well known equations of the
mazer in that case \cite{Mey97,Lof97}, namely,
\begin{equation}
\label{aeqs} i \frac{\partial C_n^{\pm}}{\partial t} = \left(
-\frac{1}{2M}\frac{\partial^2}{\partial z^2} + V_n^{\pm} \right)
C_n^{\pm},
\end{equation}
as was also stated by Abdel-Aty and Obada in Ref.~\cite{Abd02a}.
It is important to note that Eq.~(\ref{aeqs}) covers the whole $z$
axis. It describes elementary scattering processes of the
$C^{\pm}_n$ components over the potentials $V^{\pm}_n$ defined by
the atom-cavity interaction.

However, we observe that Eq.~(\ref{aeq}) does not reduce to
Eq.~(\ref{aeqs}) in the resonant case, but rather
to
\begin{equation}
i \frac{\partial C_n^{\pm}}{\partial t} = \pm \left(
-\frac{1}{2M}\frac{\partial^2}{\partial z^2} + V_n^{\pm} \right)
C_n^{\pm}
\end{equation}
which is a wrong result.

On the contrary, we may notice that, in the resonant case, our
equations (5a) and (5b) in Ref.~\cite{Bas03b} well reduces to
Eq.~(\ref{aeqs}) using $\theta = \pi/4$. Abdel-Aty's
Eq.~(\ref{aeq}) most probably contains a sign inaccuracy.
Considering that this problem should be solved on the author's own
responsibility, we should conclude that both approaches are
equivalent in the resonant case. However, we are dealing here with
the detuning effects and we must focus our attention to this
nonresonant case, for which the criticisms were written. In the
mesa mode case, it is claimed by Abdel-Aty that Eq.~(\ref{aeq}) is
equivalent to Eq.~(\ref{aeqs}) over the whole $z$ axis, arguing
that $d\theta_n/dz$ vanishes identically. This is not true and we
will be very explicit to demonstrate it. The mesa mode function
(illustrated in Fig.~\ref{mesafig}) is constant everywhere and
presents two discontinuous variations at $z=0$ and $z=L$. The
$\theta_n$ angle given by (\ref{thetan}) is thus equal to $0$ or
$\pi/2$ (according to the sign of the detuning) outside the cavity
(where $u(z) = 0$) and to $\theta_n^c$ given by $\cot 2\theta_n^c
= - \delta/2 g \sqrt{n+ 1}$ inside the cavity (where $u(z)=1$).
Therefore, $\theta_n$ is also a discontinuous function~: constant
everywhere, but with different values inside and outside the
cavity. Consequently, it is wrong to say that $d\theta_n/dz$
vanishes identically and that this factor may be eliminated
directly from Eq.~({\ref{aeq}). More precisely, $d\theta_n/dz$
vanishes everywhere \emph{except} at the entrance and at the end
of the cavity where it is infinite. This is the key point that has
been overlooked by Abdel-Aty. What happens at the cavity borders
is essential as this precisely defines the heart of the mazer
physics, namely, the study of the atom-cavity interaction for
atoms passing through the cavity. The special properties predicted
by Scully \etal \cite{Scu96} for the induced emission probability
of a photon inside the micromaser cavity by ultracold atoms
initially excited occur because the interaction between the atoms
and the field is drastically different inside and outside the
cavity, resulting in a strong effect on the atomic motion when the
cold atoms enter the cavity~\cite{Eng91}. A correct description of
the cavity borders is therefore an essential feature in the
presently investigated system and any approximation that would
elude any characteristics at these points will necessarily
profoundly affect the predictions about the system, resulting in
possible wrong results.

\begin{figure}
\begin{center}
\noindent\mbox{\includegraphics[width=7cm, bb= 90 535 515
800]{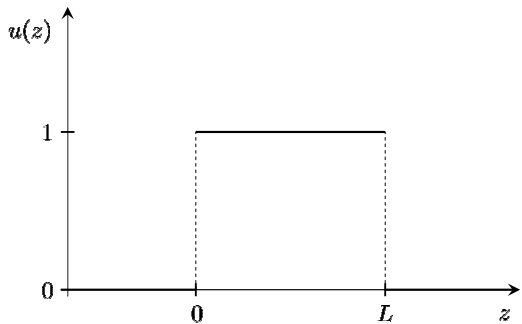}}
\end{center}
\caption{The mesa mode function.} \label{mesafig}
\end{figure}

In the mesa mode case, Eq.~(\ref{aeq}) contains two singularities
at the cavity borders that are removed in Eq.~(\ref{aeqs}). That
means that both systems of equations \emph{are not} equivalent.
They are equivalent everywhere, except at $z=0$ and $z=L$.
Following Abdel-Aty's approach, the correct way to solve the
Schr\"odinger equation would consist in considering directly
Eq.~(\ref{aeq}). However, as we just showed it now, this approach
leads to singularities in the mesa mode case. This points out the
limitations of the method and questions the validity of the
treatment. It is easy to understand the origins of these
limitations. In Abdel-Aty's approach, the
Hamiltonian~(\ref{Hamiltonian}) is diagonalized in the local
Hilbert space $\mathcal{E}_A \otimes \mathcal{E}_R$ and the atomic
centre-of-mass motion is supposed to move according to the spatial
variations of the $z$-dependent energy levels (as given by Eq.~(4)
in Abdel-Aty's Comment). It is well known that this approach is
restricted in the framework of the \emph{adiabatic approximation}.
In presence of a detuning, its validity requires smooth varying
mode functions (so that $d\theta_n/dz$ does not contain any
singularity). This condition excludes longitudinal modes of closed
cavities, whose electric fields exhibit discontinuities at the
points where the atom enters and leaves the cavity (Haroche and
Raymond \cite{Har94}). The mesa mode belongs to this exclusion
category (as it is obviously not a smoothly varying function) and
this explains why Abdel-Aty's approach cannot be followed.

On the contrary, our Eqs.~(5a) and (5b) in Ref.~\cite{Bas03b} have
been derived outside any restricted scheme in the global Hilbert
space (\ref{HilbertSpace}). Their validity is extremely general
and they can be used for any mode function, any initial atomic
wave function (including plane waves), and any detuning. They
don't contain singularities and we have shown that analytical
solutions can be found in the mesa mode case. We have
redemonstrated in this Reply the correctness of these equations
and we are thus confident about all the physical results deduced
from them in our paper. It is obvious to notice that these results
differ significantly from those obtained by Abdel-Aty and Obada in
Refs.~\cite{Abd02a} and \cite{Abd02b} (compare, for example, the
divergent expressions for the reflection and transmission
coefficients of atoms by the cavity) and that all the physical
effects we have recalled at the beginning of this Reply cannot be
deduced from their papers.

The reflection and transmission coefficients presented in
Refs.~\cite{Abd02a} and \cite{Abd02b} are those obtained in the
framework of the well-known one-dimensional scattering problems
over square potentials $V^{\pm}_n$. They are deduced from
Eq.~(\ref{aeqs}), considering that this equation would describe
scattering processes of $C^{\pm}_n$ components representing the
\emph{same} wave function projections along the whole $z$ axis.
However, this assumption is \emph{only} true in the resonant case
where the dressed state bases inside and outside the cavity are
identical ($\theta_n = \pi/4$). In the nonresonant case, these
bases differ, $C^{\pm}_n$ does not represent the same wave
function projections inside and outside the cavity and the
left-hand side of Eq.~(\ref{aeqs}) becomes $z$-dependent. In
presence of a detuning, the system \emph{does not} reduce to
elementary scattering processes over potentials $V_n^+$ and
$V_n^-$ defined by the cavity~\cite{Bas03b}. The reflection and
transmission coefficients are less evident to compute. This
explains why our results, and their results are not in agreement,
why we question them and why we cannot follow Abdel-Aty's
suggestion to replace our set of equations (5a) and (5b) in
Ref.~\cite{Bas03b} by Eq.~(\ref{aeq}). We already pointed out this
problem in a separated Comment~\cite{Bas03c}, at the same time of
the publication of our paper~\cite{Bas03b}. A Reply to our Comment
has been recently addressed by these authors~\cite{Abd04a}.
However, as it contains exactly the same criticisms as those
presented in Abdel-Aty's Comment (and the same inaccuracies
connected with Eq.~(\ref{aeq})), we refute all of them for the
same reasons detailed here and we question clearly the arguments
developed therein, where the mesa mode is dramatically confused
with a constant function along the whole $z$ axis, and the
discontinuous variations of this mode are ignored. In this sense,
the arguments developed by Abdel-Aty and Obada in
Ref.~\cite{Abd04a} and in the present Abdel-Aty's Comment are
strongly inconsistent. These authors justify the basic equations
used in Refs.~\cite{Abd02a} and \cite{Abd02b} (namely,
Eq.~(\ref{aeqs})) by arguing that the mode function does not
contain any variations. However, they consider in these papers
scattering processes and reflection mechanisms of the atoms by the
cavity, although these effects are strictly related to variations
of the mode function. If the scattering potentials were constant
everywhere, obviously no scattering could occur and this would
describe a free particle problem. This is contradictory.

As a conclusion, we have shown in this Reply that all the
criticisms raised in Abdel-Aty's Comment are not founded and
cannot be considered further. We also have demonstrated precisely
what is wrong with Abdel-Aty and Obada's arguments and why the
validity of their results published in Refs.~\cite{Abd02a},
\cite{Abd02b} and~\cite{Abd04a} must seriously be questioned.
Finally, we have justified again the correctness of our equations
and their very general validity, proving that we may be confident
in the results contained in Ref.~\cite{Bas03b}.


\end{document}